# Evaporation Dynamics of Sessile Saline Microdroplets in Oil

Ruel Cedeno[1,2], Romain Grossier[1], Nadine Candoni[1], Adrian Flood[2]*, Stéphane Veesler[1]*

[1]CNRS, Aix-Marseille University, CINaM (Centre Interdisciplinaire de Nanosciences de Marseille), Campus de Luminy, Case 913, F-13288 Marseille Cedex 09, France

[2]Department of Chemical and Biomolecular Engineering, School of Energy Science and Engineering, Vidyasirimedhi Institute of Science and Technology, Rayong 21210, Thailand

**Abstract**

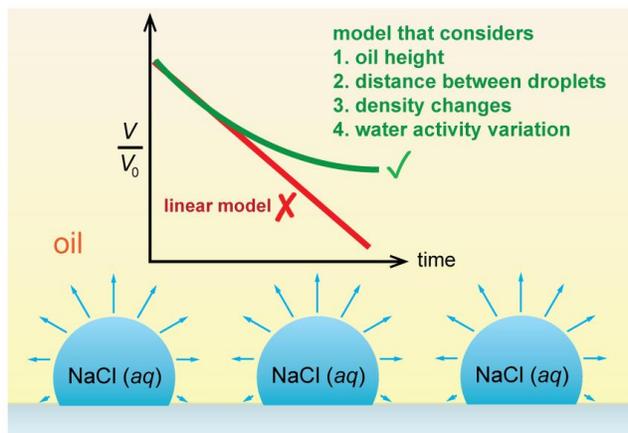

The occurrence of concentration and temperature gradients in saline microdroplets evaporating directly in air makes them unsuitable for nucleation studies where homogeneous composition is required. This can be addressed by immersing the droplet in oil under regulated humidity and reducing the volume to the picoliter range. However, the evaporation dynamics of such a system is not well understood. In this work, we present evaporation models applicable for arrays of sessile microdroplets with dissolved solute submerged in a thin layer of oil. Our model accounts for the variable diffusion distance due to the presence of the oil film separating the droplet and air, the diffusive interaction of neighboring droplets, as well as the variation of the solution density and water activity due to the evolving solute concentration. Our model shows excellent agreement with experimental data for both pure water and NaCl solution. With this model, we demonstrate that assuming a constant evaporation rate and neglecting the diffusive interactions can lead to severe inaccuracies in the measurement of droplet concentration particularly during nucleation experiments. Given the significance of droplet evaporation in a wide array of scientific and industrial applications, the models and insights presented herein would be of great value to many fields of interest.

## INTRODUCTION

Droplet evaporation on surfaces is ubiquitous in nature and plays a key role in a wide range of industrial and scientific applications[1] such as inkjet printing[2], nanostructure fabrication[3], DNA chip manufacturing[4], crystallization studies[5], biomedical diagnostics[6], as well as virus spreading[7] and testing[8]. However, this seemingly "simple" process is governed by the complex interplay of many physical phenomena such as evaporative mass transfer[9], heat conduction and convection, thermal-hydrodynamic instabilities, viscous and inertial flows, surface-tension-driven flows, contact-line pinning and depinning, buoyancy effects, and other effects.[10]

Given its complexity and practical significance, numerous experimental and theoretical investigations have been devoted to better understand the underlying physics of sessile droplet evaporation[10]. Many of these studies dealt with the evaporation of either pure liquid droplets[11-12] or those with suspended colloidal particles which can lead to the so-called "coffee-ring effect"[13-14]. However, the evaporation of droplets containing dissolved salts has been rarely investigated. For instance, Takistov et al.[15], Shin et. al.[16], Zhang et. al.[17], and Zhong et. al.[18] showed that the resulting patterns and morphologies of the dried salt droplets depend on the wettability of the surface, i.e. crystal rings would form on hydrophilic surfaces while single crystals at the center of the droplet are likely to form on hydrophobic surfaces. This suggests that surrounding salt droplets with hydrophobic liquid is a promising approach for studying nucleation inside the droplet without interaction with the hydrophobic liquid, i.e. homogeneous primary nucleation.

In the context of crystallization studies, we need to ensure spatial homogeneity of droplet temperature and composition. However, in microliter droplets, it has been shown that various internal and Marangoni flows can lead to temperature and concentration gradients[19-20]. To address this, we reduce the droplet size down to picoliter range[21] and we reduce the evaporation rate by immersing the droplet in oil under regulated humidity.[22] The oil bath also serves as a thermal buffer which minimizes temperature gradients due to evaporation. To extract nucleation parameters from such experiments[23], it is crucial to determine how the volume, and so supersaturation of microdroplets, evolve with time. In modeling the evaporation rate, Soulié et. al.[24] reported that the droplet volume varies linearly with time within the early stages of evaporation. Given that the later stages of evaporation are crucial for the analysis of nucleation, we need a model that works even for the later stages. Since we are dealing with arrays of concentrated salt microdroplets immersed in a film of oil, there are additional phenomena that need to be accounted for. First, the variable diffusion distance due to the presence of oil film separating the microdroplet and air must be taken as an additional parameter. Second, the diffusive interactions due to the presence of neighboring microdroplets must be accounted for.[25] Third, the density of the microdroplet changes as water evaporates. Fourth, the equilibrium concentration at the interface varies with time because water activity decreases as solute concentration increases (Raoult's law).[26] In this work, we derive expressions describing the evaporation dynamics that account for these four additional phenomena based on well-established mass transfer equations. We then validate our model with experimental data[27]. Moreover, we highlight that (1) surprisingly, different contact-line behavior such as constant contact angle (CCA), constant contact radius (CCR), and stick-slide (SS) leads to comparable evolution of microdroplet volume within the time of nucleation (2) failure to account for diffusive interactions between microdroplets nor the changes in colligative properties can lead to significant overestimation of their concentration.

**MODELING**

When a microdroplet is deposited onto a surface, it rapidly conforms to a quasi-equilibrium geometry with contact radius $R$ and contact angle $\theta$, which determine the droplet volume $V_d$ (**Figure 1**).

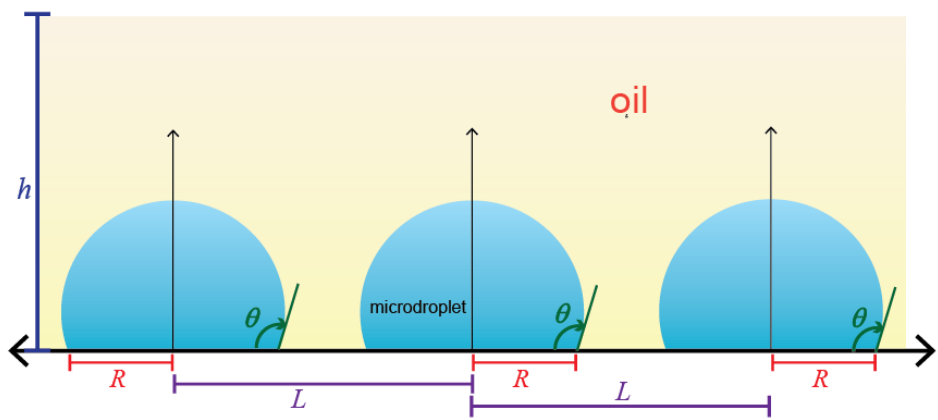

**Figure 1** Geometry of the microdroplets (modeled as a spherical cap) in a linear array with contact radius $R$, contact angle $\theta$ immersed in oil with thickness $h$. The centers of neighboring microdroplets have a distance of $L$.

## 1. Contact Line Behavior of Sessile Microdroplets

As *R* is in the micrometer range, the droplet can be assumed to be a spherical cap (see section 1 of SI, Figure A) due to the negligible gravitational effects (size is much less than the capillary length $L_c$). Thus, the droplet volume $V_d$ can be calculated as[28]

$$V_d = \pi R^3 g(\theta) \tag{1}$$

$$g(\theta) = \frac{\sin\theta\,(\cos\theta + 2)}{3(1 + \cos\theta)^2} \tag{2}$$

In the extreme case of perfectly smooth chemically homogeneous surface, the droplet maintains an equilibrium contact angle, and this is referred to as constant contact angle (CCA) mode. Consequently, during evaporation, the volume decreases due to the continuous decrease in contact radius.[28] In practice, the droplet can be pinned at some point due to surface roughness so the radius remains constant and the angle decreases due to evaporation. In the extreme case where the droplet remains pinned throughout its lifetime, we refer to this as the constant contact radius (CCR) mode. In this mode, the volume decreases due to the continuous decrease in contact angle. As experimental studies suggest,[11] real microdroplets evaporate in some mixture of CCR and CCA modes. One common observation is the occurrence of CCR mode at the beginning and once the contact angle decreases to a value less than the receding contact angle $\theta_r$, it switches to CCA mode. This combination is known as the stick-slide (SS) mode.[11] In this work, we consider all three cases (CCA, CCR, and SS models) in analyzing the experimental data.

## 2. Evaporation Rate of Sessile Droplets

In the case of diffusion-limited quasi-steady state evaporation of pure liquid droplet, Popov[26] reported an analytical expression for the mass transfer rate as follows :

$$\frac{dm}{dt} = -\pi R D M_w (c_s - c_\infty) f(\theta) \tag{3}$$

$$f(\theta) = \frac{\sin\theta}{1 + \cos\theta} + 4\int_0^\infty \frac{1 + \cosh(2\theta\delta)}{\sin(2\pi\delta)} \tanh[(\pi - \theta)\delta]\, d\delta \tag{4}$$

where *m* is the mass of the volatile species (in this case, water), *D* is the diffusivity of water in the medium, $M_w$ is the molar mass of water, $c_s$ and $c_\infty$ are the concentration of water at saturation and at a point far away from the droplet respectively (in mol/m³), $f(\theta)$ is a shape factor, and $\delta$ is an arbitrary variable of integration.

As mentioned earlier, since we are dealing with concentrated arrays of saline droplets immersed in a film of oil, there are four additional phenomena that need to be accounted for: (1) the influence of oil thickness on the evaporation rate (2) the lowering of evaporation rate due to the presence of neighboring droplets (3) the changes in droplet density as water evaporates, (4) the dependence of water activity on solute concentration.

## 2.1) Considering the influence of oil thickness on the evaporation rate

For a droplet submerged in an oil bath (R<<h), we assume an isothermal system so that temperature-dependent quantities such as solubility and diffusivity remain constant. The oil thickness is taken into consideration in our study by a factor *(1+R/2h)* introduced in equation **Erreur ! Source du renvoi introuvable.** (see section 2.1.1 of SI), leading to:

$$\frac{dm}{dt} = -\pi R D M_w (c_s - c_\infty)\left(1 + \frac{R}{2h}\right) f(\theta) \tag{5}$$

We introduce the relative humidity *RH*, defined as water concentration divided by the concentration at saturation $c_s$ (in this case, the solubility of water in oil). Note that technically, relative humidity is a vapor phase property (i.e. ratio of partial pressures). Since the liquid phase concentrations should scale proportionally to the partial pressures above them (Henry's Law), we can use *RH* to express the ratio of water concentrations in the liquid phase (i.e. oil) for simplicity. Then, we replace the transfer rate of *m* by the one of the volume *V* of pure water (see section 2.1.2 of SI), leading to:

$$\frac{dV}{dt} = -\pi R K (RH_s - RH_\infty)\left(1 + \frac{R}{2h}\right) f(\theta) \tag{6}$$

Where $K$ combines all the constant terms in $K = \frac{DM_w c_s}{\rho_w}$ with $\rho_w$ the density of pure water, $RH_s$ is the relative humidity at the droplet-oil interface (saturated). In the case of isolated droplets, we use the relative humidity at oil-air interface $RH_\infty$ in equation 6.

## 2.2) Considering the lowering of evaporation rate due to the presence of neighboring droplet

To account for the presence of neighboring droplets, we use $RH_{eff}$ (effective relative humidity) instead of $RH_\infty$ in the driving force. This is because in several studies,[29-30] the presence of neighboring droplets slow down the evaporation process relative to isolated sessile droplets. This is due to the diffusion-mediated interactions, which is a function of the relative spacing between the individual droplets. The region between the two neighboring droplets experiences an enhanced local accumulation of water, which in turn reduces the driving force for evaporation. To quantify this behavior, Hatte *et al.*[25] modeled the zone of enhanced vapor concentration as a half-cylindrical region with cross-sectional area $A_c$ (see section 2.2 of SI). Accordingly, $A_c$ is a function of the distance between the centers of the droplets *L*, the initial contact radius $R_0$, the instantaneous contact radius *R*, the initial contact angle $\theta_0$ and the instantaneous contact angle $\theta$ as

$$A_c = 4R_0 \left(L - \frac{R}{\sin\theta}\right) \sqrt{\pi\left(1 + \frac{1}{\sin\theta_0}\right)} \tag{7}$$

Based on their analysis, $RH_{eff}$ can be approximated from $RH_\infty$ using a correction factor $\epsilon$ written as :

$$\epsilon = \frac{1 - RH_{eff}}{1 - RH_\infty} = \frac{A_c}{A_c + 2\pi R_0 f(\theta_0) A \bar{L}_a} \tag{8}$$

where $R_0$ and $f(\theta_0)$ are the initial contact radius and shape factor respectively, $A$ is an empirical parameter (in the order of 1), and $\bar{L}_a$ is the average vapor accumulation length which depends on the initial geometry and on an empirical constant $\alpha$ as

$$\bar{L}_a = \frac{\alpha R_0}{\sin \theta_0} \tag{9}$$

Note that in the original derivation of Hatte et al.[25], $\alpha = 2K\beta$ where $K, \beta$ are empirical parameters which they have shown to be $AK\beta \approx 0.5$. Since A≈1, we combined these constants for simplicity giving a lumped parameter $\alpha \approx 1$. The analysis on interacting three-droplet system can then be generalized to multiple droplet arrays by virtue of symmetry.[25]

Finally, we derive the following expression for the rate of change in droplet volume:

$$\frac{dV}{dt} = -\pi R K (RH_s - RH_{eff}) \left(1 + \frac{R}{2h}\right) f(\theta) \tag{10}$$

### *2.3) Considering the changes in droplet density as water evaporates*

To account for changes in solution density as a function of concentration, we used a linear function where $\rho_w$ is the density of pure water, $S$ is the supersaturation ratio ($S = c/c_{eq}$, where c is the concentration of salt in the solution and $c_{eq}$ its solubility), $\rho$ is the density of salt solution at $S$ and $b_1$ is a coefficient fitted from experimental data. The experimental data used in these fittings are shown in Figure B of SI (see section 2.3.1 of SI).

$$\rho = \rho_w (1 + b_1 S) \tag{11}$$

Using this relation of $\rho$, we express the droplet volume $V_d$ taking into account the presence of salt and water and we replace it in equation (1) to determine the droplet radius *R in terms of S* (see section 2.3.2 of SI):

$$R = \left(\frac{V_d}{\pi g(\theta)}\right)^{\frac{1}{3}} = \left[\frac{V(1 + c_{eq} M_{salt} S)}{(1 + b_1 S) \pi \cdot g(\theta)}\right]^{\frac{1}{3}} \tag{12}$$

where *$M_{salt}$ is* salt molar mass. Then, this expression of R is used in equation (10) to determine the rate of change in droplet volume.

### *2.4) Considering the dependence of water activity on solute concentration*

To account for the lowering of water activity as concentration increases, we model the saturated relative humidity $RH_s$ as

$$RH_s = RH_0(1 - b_2 S) \tag{13}$$

where $RH_0$ is the relative humidity of air that is in an equilibrium state with pure water (equal to 1), $b_2$ is the coefficient of vapor pressure lowering fitted from experimental data[31] (see section 2.4 of SI).

## 3. Models for Contact Line Behavior

We can incorporate the contact-line behavior by modeling the behavior of the contact angle $\theta$, using the time derivative of contact angle as a function of time. The simplest case is the constant contact angle mode (CCA) in which

$$\frac{d\theta}{dt} = 0 \tag{14}$$

For constant contact radius (CCR) mode, the change in contact angle with time can be obtained by taking the time derivative of $V = f(\theta, R)$ while treating $R$ as constant (see section 1 Figure A of SI). This leads to (see section 3.2. of SI)

$$\frac{d\theta}{dt} = \frac{1}{V}\frac{dV}{dt}(1 + \cos\theta)^2 g(\theta) \tag{15}$$

Therefore, the time evolution of $V$ and $\theta$ can be obtained from the numerical solution of equation **Erreur ! Source du renvoi introuvable.** through **Erreur ! Source du renvoi introuvable.** solved simultaneously with either equation **Erreur ! Source du renvoi introuvable.** for CCA and equation **Erreur ! Source du renvoi introuvable.** for CCR.

For stick-slide mode (SS), the evaporation follows CCR mode, that is, the initial contact angle $\theta_0$ decreases until it reaches the receding contact angle $\theta_r$ where it suddenly shifts to the CCA model[28]. The full SS model can be written as (see section 3.3 of SI)

$$\frac{d\theta}{dt} = \begin{cases} \frac{1}{V}\frac{dV}{dt}(1 + \cos\theta)^2 g(\theta) & \text{for } \theta_r \leq \theta \leq \theta_0 \\ 0 & \text{for } 0 < \theta < \theta_r \end{cases} \tag{16}$$

For the numerical solution of SS, the final condition of the CCR part is used as the initial condition of the CCA part.

**MATERIALS AND METHODS**

To determine the applicability of our models, we compared the experimental results of our previous works to the numerical solution of equations (5) through (13) coupled with equations (14), (15), and (16) for CCA, CCR, and SS respectively. This gives the time evolution of droplet volume and contact angle which can then be used to calculate the contact radius and droplet height. For pure water droplets we used the data of Rodriguez-Ruiz et. al[27] which tracked the evolution of contact radius and droplet height from a series of lateral images of droplets acquired using a side-view microscope (see section 4 of SI, Table S1). With simple trigonometry, the contact radius and droplet height allow calculation of contact angle and droplet volume, assuming that microdroplets are spherical caps (see section 1 of SI, Figure A). We note that the use of side-view microscope gives direct access to geometric parameters of the microdroplets. However, it only permits measurement of 3-4 droplets at a time which is inefficient for nucleation experiments.

For saline droplets, we generated arrays of sessile NaCl microdroplets on PMMA-coated glass immersed in a thin film of PDMS oil using the method described by Grossier et al.[21] The experimental setup and additional details are shown in.[22] Properties of products are described in Table S2 of SI (see section 4 of SI). To validate our models for saline droplets, we used an approach based on the analysis of gray-level pixel standard deviation[23] of axial-view droplet images (see section **5** of SI, Figure **D**). For 190 independent microdroplets, we measure three characteristic times namely the saturation time (droplet is saturated), matching time (refractive index of droplet equals that of the oil), and nucleation time. Although the use of the bottom-view microscope only gives the droplet volume and concentration at some specific time points, it allows simultaneous measurement of hundreds of droplets, which is useful for studying the stochastic nature of nucleation.

## RESULTS AND DISCUSSION

### 1. Predictions of three models (CCR, CCA, and SS) for Pure Droplets

For water droplets with no dissolved solutes, we compared the experimental geometric parameters with the predictions of three models (CCR, CCA, and SS) in Figure 2.

The experimental points in **Figure 2**a suggests that the normalized contact radius $R/R_0$ is constant until a certain time of pinning $t_p$, after which, $R/R_0$ decreases. Meanwhile, **Figure 2**b shows that the contact angle $\theta$ decreases until this threshold at $t_p$. This behavior indicates that the system undergoes a stick-slide (SS) mode, i.e. CCR followed by CCA. In our system, we found that the time of pinning $t_p$ corresponds to a contact angle of around 86° (**Figure 2**b). Thus we assume a receding contact angle of $\theta_r$=86° for our system and we use this value for the stick-slide (SS) model in equation (15). Upon comparing the experimental points with the model predictions, it is clear that the stick-slide (SS) model well captures the evolution of the microdroplet geometric parameters (i.e. contact angle, contact radius, height, and volume). Note that although the observed deviations in contact angle and droplet height (**Figure 2**b,c) could be due to instrumental limitations, the SS model in itself is an idealization. In reality, the radius and contact angle can evolve simultaneously which is more difficult to model. Interestingly, regardless of the contact-line behavior (CCR, CCA, SS), the droplet volume evolves almost identically (**Figure 2**d). This is in agreement with Stauber et. al.[12] who showed that on strongly hydrophobic surfaces, the volume evolution of the two extreme modes CCA and CCR tend to converge for contact angles $90°<\theta<180°$.

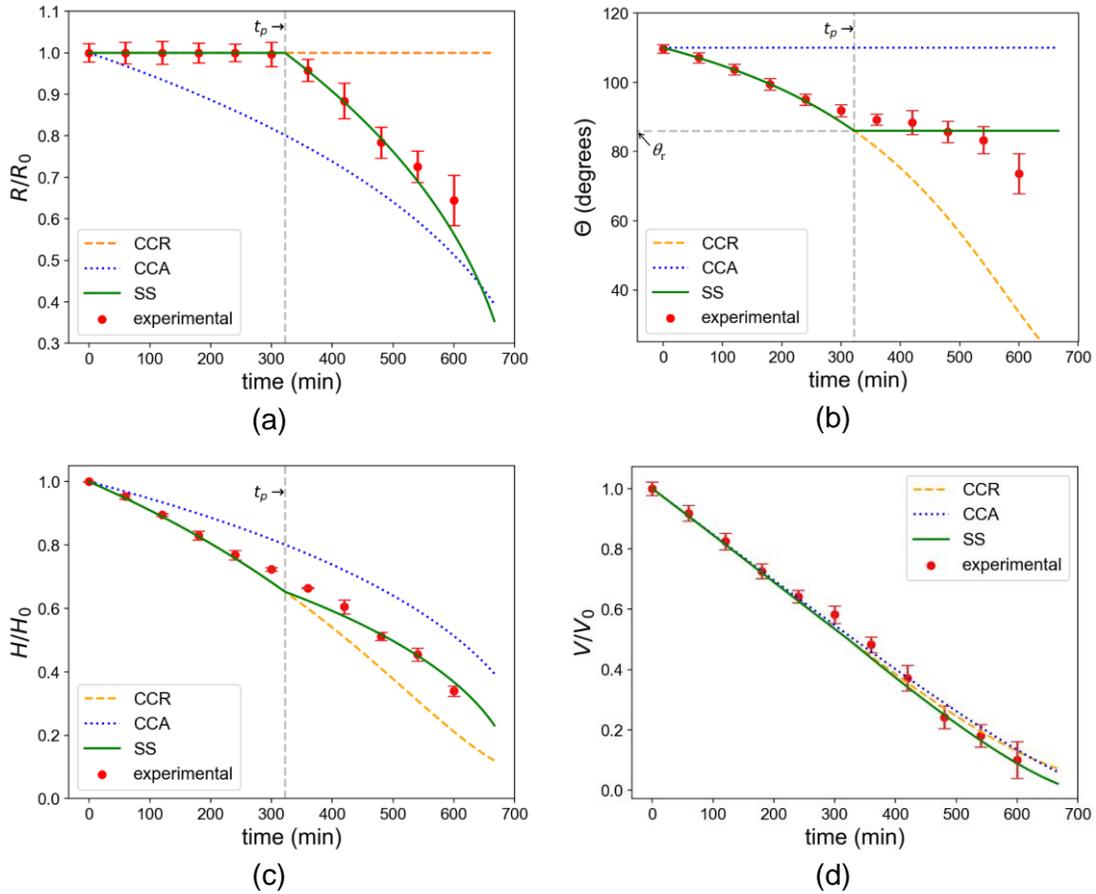

**Figure 2** Model predictions of three contact line behavior models (CCR, CCA, SS) in comparison with experimental data for pure water droplets by Rodriguez-Ruiz et. al.[27] Time evolution of (a) normalized contact radius, (b) Contact angle of the microdroplets with the substrate (c) normalized microdroplet height and (d) Volume contraction. Error bars represent standard errors based on 3 replicates and $t_p$ time of pinning.

To visualize the evolution of droplet shape, we used the numerical solution in **Figure 2** to simulate the geometry of the droplet at discrete time points. In **Figure 3**, we see that the final droplet shape is highly dependent on the contact-line behavior.

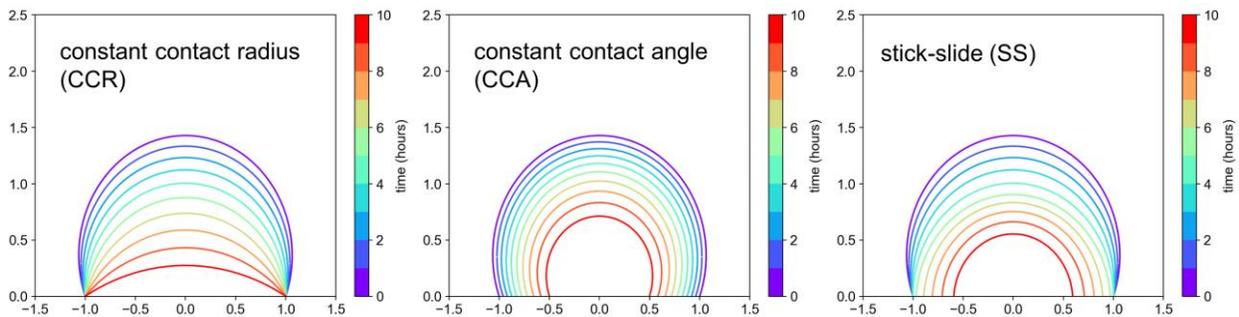

**Figure 3** Predicted evolution of microdroplet shape (pure water) for CCR, CCA, and SS models at discrete time points (every 1 hour). X,Y axis (lengths) are in terms of $R/R_0$.

Note that for the prediction of geometric parameters ($R/R_0$, $\theta$, $H/H_0$), the SS model is the most suitable. However, in the context of crystallization studies, the most important parameter to obtain from the evaporation modeling is the evolution of droplet volume on which the solution concentration depends. Thus, regardless of the droplet shape, the excellent agreement of the CCR, CCA, and SS in terms of droplet volume indicates that we can just choose one of these three contact-line behavior models to calculate the droplet concentration. In the case of saline droplets, we currently do not have an experimental value of the receding angle $\theta_r$ needed in SS model. Consequently, we use CCA to describe the evaporation rate as this is the simplest case mathematically.

## 2. The CCA model for Saline Microdroplets

Using CCA model, we thus extend our analysis to microdroplets containing dissolved salt (NaCl). As mentioned, this is based on bottom-view images from an inverted optical microscope which allow us to measure experimental points corresponding to the time at which the solution is saturated ($S=1$) and the time at which the refractive index of the droplet matches that of the oil ($S=1.395$) details are shown in reference[22] and Table S3, section 6 of SI. In **Figure 4**, we see that the CCA model is able to predict the two experimental points with excellent accuracy. Recall that in our CCA model derivation, we incorporated four important modifications to the well-established mass transfer equations. Thus, it would be interesting to see how each model modification affect the model predictions. In **Figure 4**, we see that neglecting the oil height correction can slightly overestimate the predicted volume. This is because without the oil height parameter, the droplet is considered to evaporate in an infinite medium of oil thereby hindering evaporation. Without density correction, the evaporation rate is significantly misestimated because the volume occupied by the NaCl in the droplet is not accounted which then affects the surface area to volume ratio. Remarkably, failure to correct for the relative humidity (due to the presence of neighboring droplet) and the changes in water activity (Raoult's law) led to a drastic overestimation of evaporation rate. This is because both cases directly affect the driving force for evaporation. The relationship between the effective relative humidity $RH_{eff}$ and the prevailing humidity above the oil $RH_\infty$ is shown in Figure E in section 6.1 of SI. Moreover, notice that neglecting the effect of colligative properties results in a linear decrease of droplet volume which is consistent with what is observed in pure droplets.

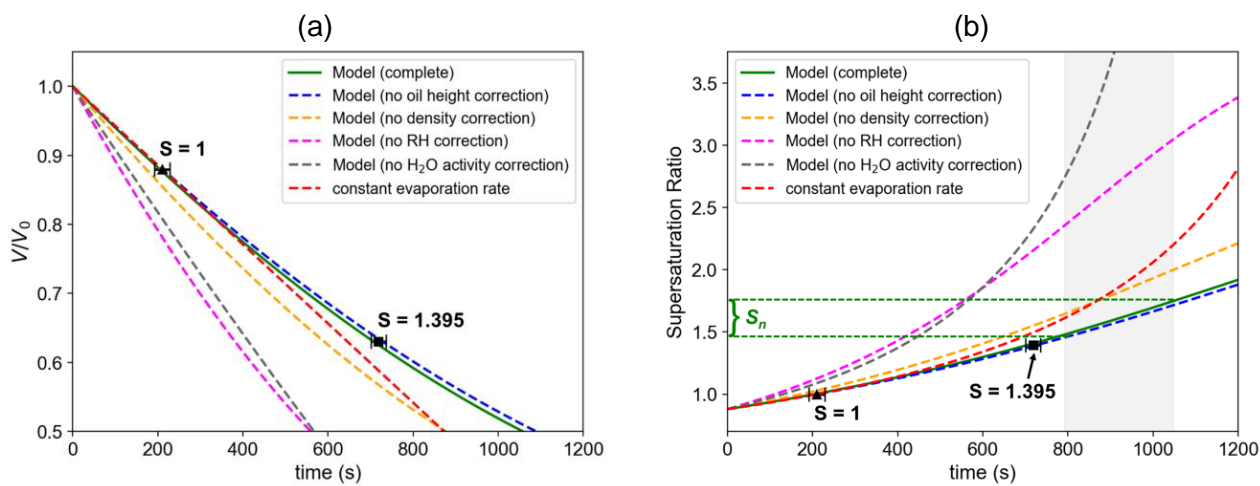

**Figure 4** Model predictions (CCA) for saline microdroplets ($V_0$ = 67 pL and $S_0$ = 0.88) in terms of (a) droplet volume and (b) supersaturation ratio in comparison with experimental data. The error bars at saturation time (S=1) and matching time (S=1.395) represent the standard deviation of the distribution of data points (190 droplets).

To verify whether the saline droplets have a homogeneous composition throughout the evaporation process, we plotted the Peclet number as a function of time (see Section 6.2. of SI and Figure F) and we found that the maximum *Pe* is in the order of $10^{-4}$ suggesting a uniform droplet concentration.

In the context of crystallization studies, droplets are not expected to nucleate at the same time even though they have identical concentration due to the stochastic nature of nucleation. Our results demonstrate this with nucleation events spanning from 800 s to 1050 s (the grey area in **Figure 4**b). In principle, these nucleation times can be used to estimate the interfacial surface energy of NaCl-water if we know the supersaturation ratio at nucleation $S_n$. To do this, several reports assumed a linear evaporation rate (neglecting changes in water activity) to calculate the droplet concentration as a function of time[32-34]. Here we highlight that this approximation can lead to inaccurate values of droplet concentration particularly in later stages where nucleation occurs. For instance, using our model, the supersaturation at nucleation $S_n$ ranges from S = 1.50 to 1.75 (**Figure 4**b). This is consistent with the results of Desarnaud et. al.[35] who showed a metastability limit of S = 1.60 for NaCl-water system using microcapillary experiments. However, if we assume a constant evaporation rate by extrapolating $t = 0$ and $t =$ saturation time, the predicted range of $S_n$ would be up to 40% higher (ranging from 1.75 to 2.50). This discrepancy would have a huge consequence particularly in crystallization studies. To illustrate this, we plot the cumulative probability distribution as a function of supersaturation at nucleation $S_n$ in **Figure 5**. The constant evaporation rate assumption clearly overestimates $S_n$ resulting in unreasonably large values of supersaturation. Furthermore, if diffusive interactions and changes in water activity were not accounted for, much larger deviations could be obtained. All of these can lead to inaccurate values of nucleation kinetic parameters. Thus, we highlight the need for accurate modeling of evaporation rate of sessile droplets in the context of nucleation studies.

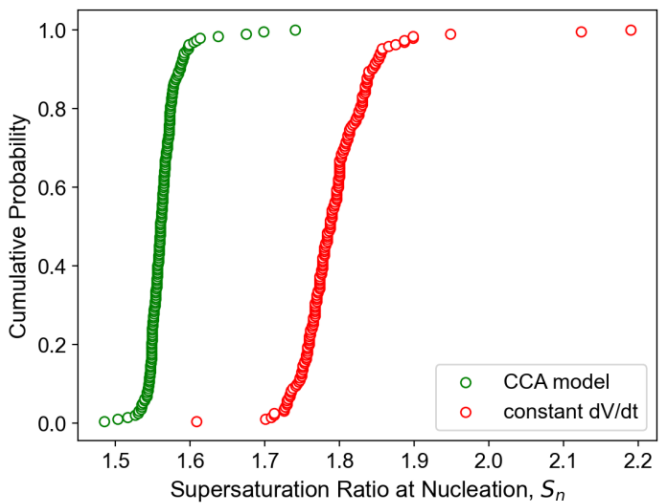

**Figure 5** Cumulative probability distribution of supersaturation ratio at nucleation $S_n$ based on two evaporation models.

## CONCLUSION

In this work, we studied the evaporation dynamics of sessile saline picoliter droplets in oil until crystallization. Starting from well-established mass transfer equations for pure sessile droplets evaporating in air, we derived new expressions applicable for droplets with dissolved solute submerged in a thin layer of oil. Our model accounts for the additional complexity due to (i) variable diffusion distance due to the presence of oil (ii) diffusive interactions due to the presence of neighboring droplets (iii) density change as concentration increases (iv) water activity change as a function of concentration. By comparing our model predictions to experimental data, we showed that different contact-line behavior (CCR, CCA, or SS) results in almost identical evolution of droplet volume especially within the time scale relevant to crystallization studies. With this information, we analyzed the evaporation rate of saline droplets using the CCA model and using NaCl-water as a model system, we demonstrated for the first time that assuming a constant evaporation rate as well as neglecting the diffusive interactions between droplets can lead to severe discrepancies in the measurement of droplet concentration particularly during nucleation. This indicates that crystallization studies in literature that had used this assumption may be subject to large errors (In the example presented here, 40%). With our model, we can accurately determine the time evolution of droplet concentration which is important in quantifying crystallization kinetics. Moreover, given the importance of evaporation dynamics in a wide array of scientific and practical applications, our models and new insights presented herein would be of great value to many fields of interest.

## ACKNOWLEDGEMENTS

R. Cedeno acknowledges the financial support of Vidyasirimedhi Institute of Science and Technology (VISTEC) and the Eiffel Excellence Scholarship (N°P744524E) granted by the French Government.

## SUPPORTING INFORMATION

Contact Line Behavior of Sessile Microdroplets
Evaporation Rate of Sessile Droplets
Models for Contact Line Behavior
Parameters and properties
Measurement of Characteristic Time Points in Saline Droplets
The CCA model for Saline Microdroplets

This information is available free of charge via the Internet at http://pubs.acs.org/

# Supporting Information for:

# Evaporation Dynamics of Sessile Saline Microdroplets in Oil

Ruel Cedeno[1,2], Romain Grossier[1], Nadine Candoni[1], Adrian Flood[2*], Stéphane Veesler[1*]

[1]CNRS, Aix-Marseille University, CINaM (Centre Interdisciplinaire de Nanosciences de Marseille), Campus de Luminy, Case 913, F-13288 Marseille Cedex 09, France

[2]Department of Chemical and Biomolecular Engineering, School of Energy Science and Engineering, Vidyasirimedhi Institute of Science and Technology, Rayong 21210, Thailand

## 1. Contact Line Behavior of Sessile Microdroplets

When a droplet is deposited onto a surface, it rapidly conforms to a quasi-equilibrium geometry with contact radius $R$, and contact angle $\theta$, which determine the droplet volume $V_d$. The shape of the droplet is either spherical or flattened, depending on the value of $R$ compared to the capillary length $L_c$ which characterizes the ratio of the interfacial energy between the droplet and the medium $\gamma_{(droplet/medium)}$ to gravitational effects. $L_c$ can be calculated as

$$L_c = \sqrt{\frac{\gamma_{(droplet/medium)}}{\Delta\rho \times g}} \quad (S1)$$

where $\Delta\rho$ is the density difference between the solution and the surrounding medium and $g$ is the gravitational acceleration. In our case, the droplet is either pure water or saline solution and the medium is PDMS oil. If the droplet size is much less than $L_c$, then the droplet assumes a spherical cap geometry. For the PDMS-water system[1], the capillary length is in the millimeter range. Since $R$ is in the micrometer range, (much smaller than $L_c$), the gravity is negligible compared to the interfacial energy between droplet and oil and so the droplets can be assumed to be a spherical cap. Thus, the droplet volume $V_d$ can be calculated as[2]

$$V_d = \pi R^3 g(\theta) \text{ with } g(\theta) = \frac{\sin\theta(\cos\theta + 2)}{3(1+\cos\theta)^2} \quad (S2)$$

In the following section, we derive expressions for the diffusion-controlled evaporation of saline microdroplet with contact radius $R$ and constant contact angle $\theta$ immersed in a PDMS oil bath with thickness $h$. The different cases ($\theta$>90º, $\theta$=90º, $\theta$<90º) are shown in **Figure A**. Recall that we define $r$ as the radial distance from the center of the equivalent spherical cap at an angle of $\phi$ with the equatorial line.

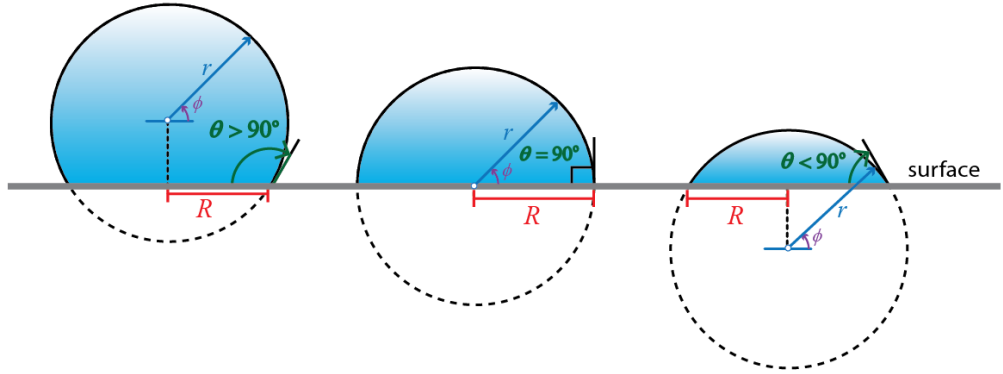

**Figure A**. Illustration of microdroplet showing the equivalent spherical cap at different values of contact angle $\theta$.

For simplicity, we will first consider the case where $\theta=90°$ (hemispherical droplet) which exhibits uniform evaporation flux over the surface area. Later on, we will incorporate a widely-used shape factor[3-4] denoted as $f(\theta)$ to obtain a general expression for any value of $\theta$.

## 2. Evaporation Rate of Sessile Droplets

### 2.1 Influence of oil thickness on the evaporation rate

*2.1.1. Introduction of the factor (1+R/2h) (leading to equation 5 in the main text)*
Since the microdroplet is submerged in an oil bath (R<<h), we assume an isothermal system so that temperature-dependent quantities such as solubility and diffusivity remain constant. With the continuity equation in spherical coordinates, the molar flux of water vapor $N(r)$ as a function of radial distance $r$ is

$$\frac{1}{r^2}\frac{d}{dr}(r^2 N) = 0 \Rightarrow N(r) = \frac{C_1}{r^2} \tag{S3}$$

where $C_1$ is a constant of integration that will be evaluated later. Assuming negligible convective transport, Fick's equation can be simplified as

$$N = -D\frac{dc}{dr} \tag{S4}$$

where $D$ is the diffusivity of water in oil and $c$ is the molar concentration of water. Combining equations (S3) and (S4),

$$\frac{dc}{dr} = -\frac{C_1}{D}\left(\frac{1}{r^2}\right) \tag{S5}$$

Since the diffusion distance varies at any angle (with respect to the horizontal), the radial distance from the droplet center to the oil-air interface is $r = \frac{h}{\sin\phi}$. To facilitate integration, we express the boundary conditions in terms of $R$. We can write $\frac{h}{\sin\phi} = R + \frac{h - R\sin\phi}{\sin\phi}$. Given that $h \gg R\sin\phi$, we can approximate $\frac{h}{\sin\phi} \approx R + \frac{h}{\sin\phi}$. Integrating equation (S5) with boundary conditions $c(R) = c_s$ and $c\left(R + \frac{h}{\sin\phi}\right) = c_\infty$, we obtain

$$\int_{c_s}^{c_\infty} dc = -\frac{C_1}{D}\int_{r=R}^{r=R+\frac{h}{\sin\phi}}\left(\frac{1}{r^2}\right)dr \Rightarrow C_1 = D(c_\infty - c_s)\left(\frac{1}{R} - \frac{1}{R + \frac{h}{\sin\phi}}\right)^{-1} \tag{S6}$$

Combining equations (S3), and (S6), we can write the molar flux as

$$N(r,\phi) = D(c_\infty - c_s)\left(\frac{1}{R} - \frac{1}{R + \frac{h}{\sin\phi}}\right)^{-1}\left(\frac{1}{r^2}\right) \tag{S7}$$

Now, we can express the rate of change in droplet volume as the mass flux of water vapor integrated over the droplet surface area A.

$$\frac{dm}{dt} = \int^A M_w N(r,\phi)dA \tag{S8}$$

The differential surface area $dA$ can be written as a function of the differential angle $d\phi$ as

$$dA = (2\pi r \cos\phi)(rd\phi) \tag{S9}$$

Combining equations (S7), (S8) and (S9) and integrating $\phi$ from 0 to $\pi/2$ (because we consider the case of hemispherical droplet where $\theta = \pi/2$), we get

$$\frac{dm}{dt} = \int_0^{\frac{\pi}{2}} M_w D(c_\infty - c_s)\left(\frac{1}{R} - \frac{1}{R + \frac{h}{\sin\phi}}\right)^{-1}\left(\frac{1}{r^2}\right)(2\pi r \cos\phi)(rd\phi) \tag{S10}$$

$$\frac{dm}{dt} = -(2\pi R)DM_w \left(c_s - c_\infty\right)\left(1 + \frac{R}{2h}\right) \quad \text{(S11)}$$

Note that this is similar to that of Popov[5] for pure droplets directly evaporating in air. By comparison, if we substitute θ = π/2 in the shape factor expression (equation 4 in the main text), we obtain *f(θ)* = 2 (via numerical integration), *i.e.*

$$f\left(\frac{\pi}{2}\right) = \frac{\sin\left(\frac{\pi}{2}\right)}{1 + \cos\left(\frac{\pi}{2}\right)} + 4\int_0^\infty \frac{1 + \cosh(2(0.5\pi)\varepsilon)}{\sin(2\pi\varepsilon)} \tanh\left[\left(\pi - \frac{\pi}{2}\right)\varepsilon\right] d\varepsilon = 2 \quad \text{(S12)}$$

Furthermore, note that $\left(1 + \frac{R}{2h}\right) \approx 1$ since R<<h. Thus, we incorporate the shape factor *f(θ)* for any contact angle *θ* as

$$\frac{dm}{dt} = -\pi R DM_w \left(c_s - c_\infty\right)\left(1 + \frac{R}{2h}\right) f(\theta) \quad \text{(S13)}$$

*2.1.2. Introduction of the relative humidity (leading to equation 6 in the main text)*

The relative humidity *RH* is defined as water vapor concentration divided by the concentration at saturation $c_s$ (in this case, the solubility of water in oil). Thus, we can write

$$\left(c_s - c_\infty\right) = c_s \left(RH_s - RH_\infty\right) \quad \text{(S14)}$$

where $RH_s$ and $RH_\infty$ are the relative humidity at the droplet-oil interface (saturated) and oil-air interface, respectively. For pure water droplets, $RH_s$ is always equal to 1. As a result, equation (S13) can be written as

$$\frac{dm}{dt} = -\pi R DM_w c_s \left(RH_s - RH_\infty\right)\left(1 + \frac{R}{2h}\right) f(\theta) \quad \text{(S15)}$$

Note that *m* is the mass of the volatile component (in this case, water). Using the definition of density, we can write *m* = $\rho_w V$ where $\rho_w$ and *V* are the density and volume of pure water respectively. Since $\rho_w$ is constant, $\frac{dm}{dt} = \rho_w \frac{dV}{dt}$. We can then combine the constant terms as $K = \frac{DM_w c_s}{\rho_w}$. Thus, equation (S15) can be re-written as

$$\frac{dV}{dt} = -\pi R K \left(RH_s - RH_\infty\right)\left(1 + \frac{R}{2h}\right)f(\theta) \tag{S16}$$

Note that this is valid for isolated droplets (i.e., no neighbors).

## 2.2. Considering the presence of neighboring droplet

Several studies,[6-7] have shown that the presence of neighboring droplets slow down the evaporation process relative to isolated sessile droplets due to the contribution of the neighboring droplets to the local relative humidity. To account for this behavior, we adapt the theoretical model of Hatte et al.[8] In simple terms, the effective relative humidity $RH_{eff}$ is approximated from the prevailing relative humidity at the oil-air interface $RH_\infty$ using a correction factor $\epsilon$ defined as

$$\epsilon = \frac{1 - RH_{eff}}{1 - RH_\infty} = \frac{A_c}{2\pi R_0 f(\theta_0) A \bar{L}_a + A_c} \tag{S17}$$

where $A_c$ is the surface area of the vapor field, $R_0$ and $f(\theta_0)$ are the initial contact radius and shape factor respectively, and $\bar{L}_a$ is the average vapor accumulation length (Refer to equation 8.8 of Hatte et al.[8]). Accordingly, $A_c$ is the cross-sectional area of the half-cylindrical region surrounding the microdroplets with enhanced local vapor concentration. This is a function of the distance between the centers of the droplets $L$, the initial contact radius $R_0$, the instantaneous contact radius $R$, the initial contact angle $\theta_0$ and the instantaneous contact angle $\theta$ as

$$A_c = 4R_0\left(L - \frac{R}{\sin\theta}\right)\sqrt{\pi\left(1 + \frac{1}{\sin\theta_0}\right)} \tag{S18}$$

$$\bar{L}_a = \frac{\alpha R_0}{\sin\theta_0} \tag{S19}$$

where $\alpha$ is a constant. Note that in the original derivation of Hatte et al.[8], we substituted $\lambda = 2L - D_e$ (see their Figure 6a) and $D_e = 2R/\sin\theta$ (spherical cap geometry). We also let $\alpha = 2K\beta$ where $K$ and $\beta$ are empirical parameters which have been shown to follow $K\beta \approx 0.5$. For simplicity, we combined this giving a single parameter ($\alpha \approx 1$).
Thus, for droplets with neighbors, we replace $RH_\infty$ by $RH_{eff}$ in equation (S16) which leads to

$$\frac{dV}{dt} = -\pi RK(RH_s - RH_{eff})\left(1 + \frac{R}{2h}\right)f(\theta) \qquad (S20)$$

## 2.3 Considering the changes in droplet density as water evaporates

### 2.3.1. Determination of the density of salt solution

Note that we defined $V$ as the volume of pure water (the volatile component) and $R$ as the radius of the entire droplet. However, the total volume of the droplet $V_d$ is a function of the volume occupied by both water and salt ions. To relate the volume of pure water $V$ to the droplet volume $V_d$, we employ experimental data on the solution density change as a function of NaCl supersaturation ratio ($S = c/c_{eq}$, where c is the concentration of salt in the solution and $c_{eq}$ its solubility) as shown in **Figure B**. We then use a simple linear function with $b_1$ (slope) as the dimensionless coefficient of density increase relating the density of pure water $\rho_w$ and the density $\rho$ at any $S$.

$$\rho = \rho_w(1 + b_1 S) \qquad (S21)$$

### 2.3.2. Determination of droplet radius

Given that the droplet mass is the sum of water mass and NaCl mass ($m_d = m_w + m_{NaCl}$), we can write

$$\frac{m_w + m_{NaCl}}{V_d} = \frac{m_w}{V}(1 + b_1 S) \implies \frac{1 + \left(\frac{m_{NaCl}}{m_w}\right)}{V_d} = \frac{1 + b_1 S}{V} \qquad (S22)$$

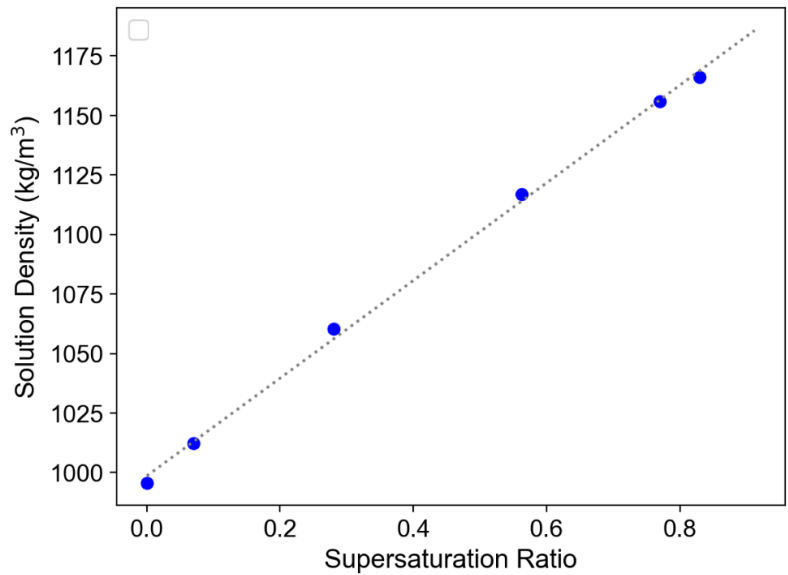

**Figure B**. Variation of aqueous NaCl density as a function of supersaturation ratio.[9] The regression line is $y = 998(1+0.205x)$ with $R^2 = 0.9984$.

We can express $m_{NaCl}/m_w$ in terms of $S$ using the solubility of NaCl in water $c_{eq}$ (in mol/kg water) and NaCl molar mass $M_{NaCl}$ (kg/mol)

$$\frac{m_{NaCl}}{m_w} = c_{eq} M_{NaCl} S \qquad (S23)$$

Thus, the droplet volume $V_d$ is related to the volume of pure water $V$ as

$$V_d = \left(\frac{1 + c_{eq} M_{NaCl} S}{1 + b_1 S}\right) V \qquad (S24)$$

Observe that for pure droplet ($S=0$), $V_d = V$. We can now express the droplet radius $R$ in terms of $V$ using the equation for the volume of spherical cap along with the density changes.

$$R = \left(\frac{V_d}{\pi g(\theta)}\right)^{\frac{1}{3}} = \left[\frac{V(1 + c_{eq} M_{NaCl} S)}{(1 + b_1 S)\pi \cdot g(\theta)}\right]^{\frac{1}{3}} \text{ with } g(\theta) = \frac{\sin\theta (\cos\theta + 2)}{3(1 + \cos\theta)^2} \qquad (S25)$$

This expression for $R$ will be used in equation (S20).

### 2.4. Dependence of water activity on solute concentration

To account for the change in water activity due to the presence of salt, we express the decrease in water activity as a linear function with slope $b_2$ fitted from experimental data of An et al, as shown in **Figure C**.[10]

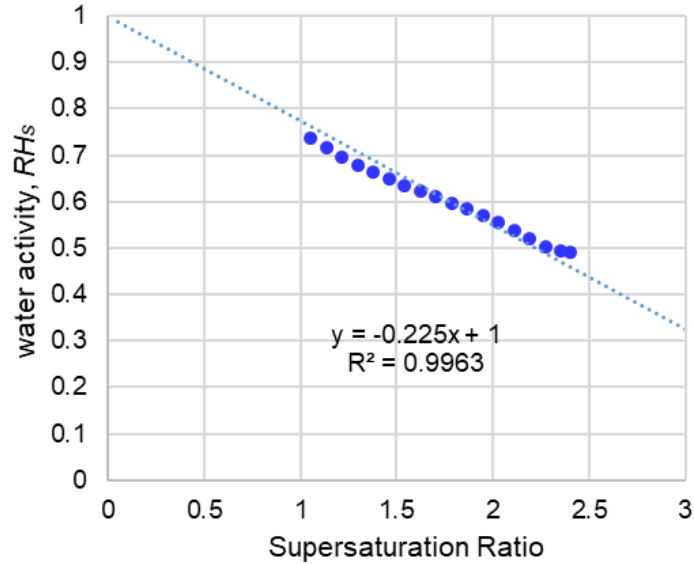

**Figure C**. Variation of water activity (numerically equal to the equilibrium relative humidity, $RH_s$) as a function of supersaturation ratio. The data were taken from Table 6 of An et al.[10]

Thus, in equation (S20), the saturation relative humidity $RH_s$ is expressed as

$$RH_s = 1 - b_2 S \qquad (S26)$$

where $b_2$ is the coefficient of vapor pressure lowering fitted from experimental data of An et al.[10] Since the total mass of the salt is constant, we can write $S_0 V_0 = SV$ so all equations containing $S$ can be expressed in terms of $V$.

## 3. Models for Contact Line Behavior

The contact line behavior (how the contact radius and contact angle evolve with time) generally depends on the nature of the surface where the sessile microdroplet is situated. In the extreme case of perfectly smooth chemically homogeneous surface, the droplet maintains an equilibrium contact angle, and this is referred to as constant contact angle (CCA) mode. Consequently, the volume decreases due to the continuous decrease in contact radius.[2] In practice, the droplet will be pinned due to surface roughness so the radius remains constant at some point. In the extreme case where the droplet remains pinned throughout its lifetime, we refer to this as the constant contact radius (CCR) mode. In this mode, the volume decreases due to the continuous decrease in contact angle. As experimental studies suggest,[11] real droplets evaporate in some mixture of CCR and CCA modes. One common observation is the occurrence of CCR mode at the beginning and once the contact angle decreases to a value less than the receding contact angle $\theta_r$, it switches to CCA mode. This combination is known as the stick-slide (SS) mode.[11] In this work, we consider all three cases (CCA, CCR, and SS models) in analyzing the experimental data.

Mathematically, we can then incorporate the contact-line behavior by modeling the behavior of the contact angle $\theta$.

### 3.1. For constant contact angle mode (CCA)
For constant contact angle mode (CCA), the change in contact angle with time is simply,

$$\frac{d\theta}{dt} = 0 \qquad (S27)$$

### 3.2. For constant contact radius mode (CCR)
For constant contact radius mode (CCR), the change in contact angle with time can be obtained by taking the derivative of $V = f(\theta, R)$ where $R$ is constant (see **Figure A**)

$$V = \pi R^3 g(\theta) \Rightarrow \frac{dV}{dt} = \pi R^3 \frac{d}{dt}[g(\theta)] \text{ with } g(\theta) = \frac{\sin\theta \, (\cos\theta + 2)}{3(1 + \cos\theta)^2} \qquad (S28)$$

$$\frac{d}{dt}[g(\theta)] = \frac{d}{dt}\left(\frac{\sin\theta(\cos\theta+2)}{3(1+\cos\theta)^2}\right) = \frac{1}{(1+\cos\theta)^2}\frac{d\theta}{dt} \qquad (S29)$$

Combining equations (S28) and (S29), we can obtain the change in contact angle as

$$\frac{d\theta}{dt} = \frac{1}{V}\frac{dV}{dt}(1+\cos\theta)^2 g(\theta) \qquad (S30)$$

### 3.3. For stick-slide mode (SS)

For stick-slide mode (SS), the evaporation follows CCR mode, that is, the initial contact angle $\theta_0$ decreases until it reaches the receding contact angle $\theta_r$ where it suddenly shifts to the CCA model[2]. Then, the full SS model can be written as

$$\frac{d\theta}{dt} = \begin{cases} \frac{1}{V}\frac{dV}{dt}(1+\cos\theta)^2 g(\theta) & \text{for } \theta_r \leq \theta \leq \theta_0 \\ 0 & \text{for } 0 < \theta < \theta_r \end{cases} \qquad (S31)$$

## 4. Parameters and properties

**Table S1** *Numerical values used as input in the evaporation model of pure water droplets*

| Experimental Parameter | Symbol | Value | Unit |
|---|---|---|---|
| initial radius | $R_0$ | 25.7 | µm |
| initial contact angle | $\theta_0$ | 110 | degrees |
| initial volume | $V_0$ | 64.6 | pL |
| receding contact angle (for SS) | $\theta_r$ | 86 | degrees |
| oil height | $h$ | 0.40 | mm |
| ambient temperature | $T$ | 298 | K |
| relative humidity at evaporation step | $RH_\infty$ | 60 | % |
| distance between droplet centers | $L$ | 65 | µm |
| **Literature Data** | | | |
| solubility of water in paraffin oil[12] | $c_s$ | 2.95 | mol/m$^3$ |
| diffusivity of water in paraffin oil[13] | $D$ | 8.5×10$^{-10}$ | m$^2$/s |
| density of pure water[9] | $\rho_w$ | 997 | kg/m$^3$ |

*Table S2. Properties of products*

| Product | Supplier | Properties |
|---|---|---|
| Sodium chloride, NaCl | R.P Normapur ® | Purity = 99.5% <br> Refractive index = 1.5442 |
| Polymethylmethacrylate, PMMA | ALLRESIST GmbH | Molecular weight= 950,000 g/mol <br> Refractive index = 1.395 |
| Polydimethylsiloxane, PDMS oil | Alfa Aesar | Molecular weight = 1250 g/mol <br> Viscosity = 10 cSt <br> Refractive index = 1.3990 |
| Ultrapure water | via Milli-Q Purifier | resistivity = 18.2 MΩ·cm <br> TOC value < 5 ppb |

# 5. Measurement of Characteristic Time Points in Saline Droplets

To validate our models for saline droplets, we generated arrays of sessile saline microdroplets on PMMA-coated glass immersed in a thin film of PDMS oil using the method described by Grossier et al.[14]

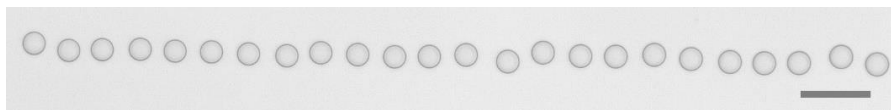

**Figure D**. Image of a typical microdroplet array (scale bar = 200 µm)

On 190 independent microdroplets, we measured three characteristic times namely the saturation time (time when the microdroplet is saturated, S=1), the matching time (time when the refractive index of the droplet matches that of the PDMS oil, S = 1.395) and the nucleation time. Our approach is to use image analysis to determine these three points as demonstrated in our previous work [Faraday 2021]. Briefly, the standard deviation of the gray-level pixel histogram (denoted as σ, a function of refractive index difference) corresponding to the region surrounding the microdroplet image (axial view) is used as an indicator of droplet concentration.

# 6. The CCA model for Saline Microdroplets

**Table S3** Numerical values used as input in the CCA evaporation model of saline droplets

| Experimental Parameter | Symbol | Value | |
|---|---|---|---|
| initial radius | $R_0$ | 26.1 | µm |
| contact angle | $\theta$ | 110 | degrees |
| initial volume | $V_0$ | 66.93 | pL |
| radius at saturation | $R_s$ | 25 | µm |
| oil height | $h$ | 0.40 | mm |
| ambient temperature | $T$ | 298 | K |
| rel. humidity at evaporation step | $RH_\infty$ | 10 | % |
| distance between droplet centers | $L$ | 100 | µm |
| **Literature Data** | | | |
| solubility of water in PDMS oil[15] | $c_s$ | 30 | mol/m$^3$ |
| diffusivity of water in PDMS oil[16] | $D$ | 8.5×10$^{-10}$ | m$^2$/s |
| coefficient of density change[9] | $b_1$ | 0.205 | - |
| coefficient of water activity lowering[10] | $b_2$ | 0.225 | - |
| solubility of NaCl in water[17] | $c_{eq}$ | 6.14 | mol/kg |
| molar mass of NaCl | $M_{NaCl}$ | 0.0584 | kg/mol |
| diffusivity of NaCl in water[18] | $D_i$ | 1.47×10$^{-9}$ | m$^2$/s |
| density of pure water[9] | $\rho_w$ | 997 | kg/m$^3$ |

## 6.1. The effective relative humidity $RH_{eff}$

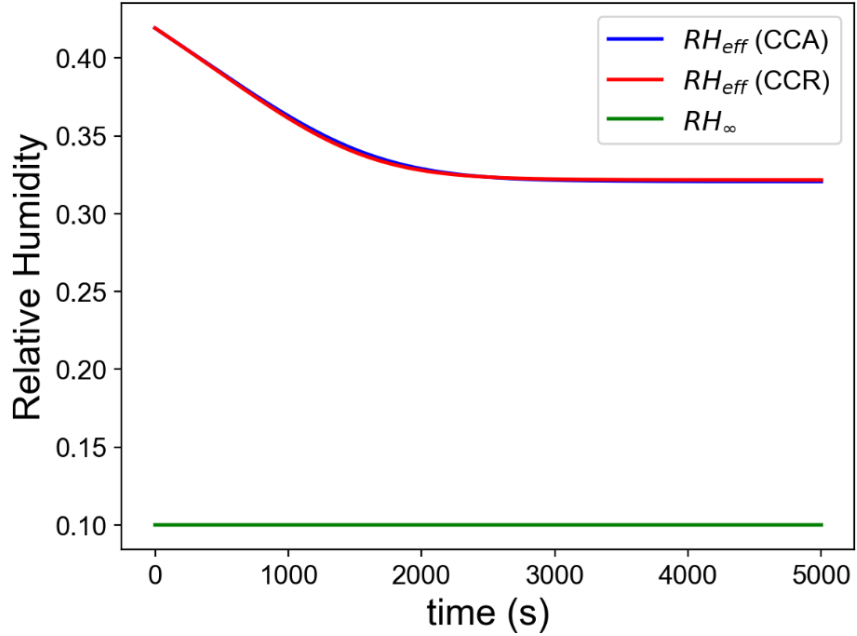

**Figure E.** Evolution of the effective relative humidity $RH_{eff}$ for CCA and CCR against the humidity above the oil $RH_\infty$. $RH_{eff}$ decreases and then reaches equilibrium as the droplets become very small.

## 6.2. Homogeneity of Droplet Concentration

In the equations (S31), we approximate that the temperature and concentration are essentially homogeneous. The homogeneity of droplet composition is characterized by Peclet number $Pe$, which is the ratio of convective mass transfer to diffusive mass transfer[19]. It is expressed as[20]

$$Pe = \frac{2R\kappa}{D_i} \tag{32}$$

Where $\kappa$ is the evaporation flux (volume loss $dV/dt$ per unit area $A$), $R$ is the droplet radius and $D_i$ is the diffusion coefficient of the solute in the droplet. If Pe < 1, the diffusion rate of the solute is fast enough to avoid a considerable enrichment at the receding surface and thus the system maintains a homogeneous composition. In our experiments, $Pe$ is in the order of $10^{-4}$ (Figure F), thus we can treat the microdroplets as homogeneous solution (with negligible concentration gradient).

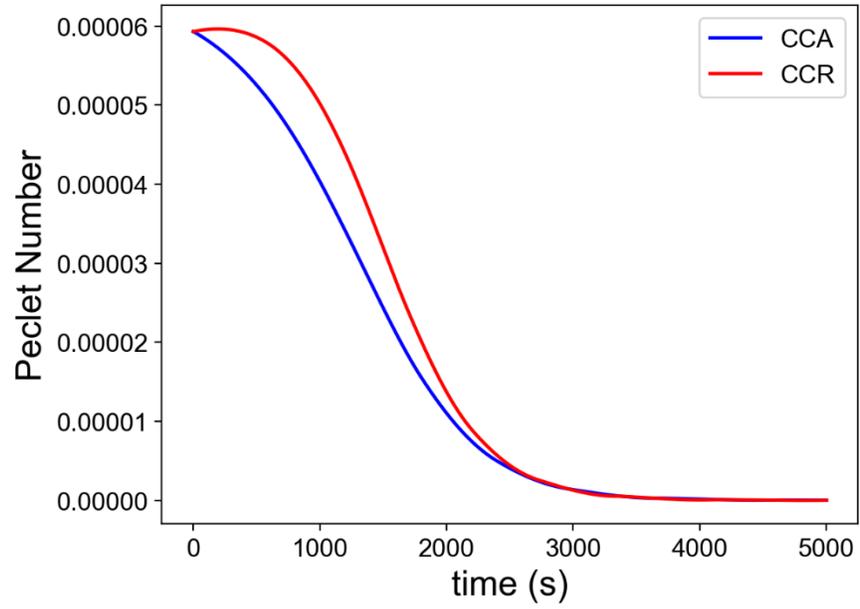

**Figure F.** Evolution of Peclet number for CCA and CCR models. (If Peclet number << 1, the microdroplet is considered to have homogeneous composition)